\documentclass[aps,prd,secnumarabic,amssymb,nobibnotes, amsmath,twocolumn,nofootinbib,superscriptaddress,floatfix,preprintnumbers]{revtex4-1}
\usepackage{docs}
\usepackage{bm}
\usepackage{ulem}
\usepackage[colorlinks=true,linkcolor=blue]{hyperref}%
\usepackage{comment}
\usepackage{graphicx}%
\usepackage{dcolumn}%

\usepackage{url}
\usepackage{gensymb}
\usepackage{tabularx}
\usepackage{xcolor}
\usepackage{multirow}

\newcommand{\Fermi}{{\it Fermi}}
\newcommand{\Chandra}{{\it Chandra}}
\newcommand{\Gaia}{{\it Gaia}}

\begin{document}

\preprint{LAPTH-039/20}

\title{The Galactic bulge millisecond pulsars shining in X rays: A $\gamma$-ray perspective}%

\author{Joanna Berteaud}
\email{berteaud@lapth.cnrs.fr}
\affiliation{Univ.~Grenoble Alpes, USMB, CNRS, LAPTh, F-74940 Annecy, France}
\author{Francesca Calore}%
\email{calore@lapth.cnrs.fr}
\affiliation{Univ.~Grenoble Alpes, USMB, CNRS, LAPTh, F-74940 Annecy, France}
\author{Ma\"{i}ca Clavel}%
\affiliation{Univ.~Grenoble Alpes, CNRS, IPAG, F-38000 Grenoble, France}
\author{Pasquale Dario Serpico}%
\affiliation{Univ.~Grenoble Alpes, USMB, CNRS, LAPTh, F-74940 Annecy, France}
\author{Guillaume Dubus}%
\affiliation{Univ.~Grenoble Alpes, CNRS, IPAG, F-38000 Grenoble, France}
\author{Pierre-Olivier Petrucci}%
\affiliation{Univ.~Grenoble Alpes, CNRS, IPAG, F-38000 Grenoble, France}

\date{\today}

\begin{abstract}
If the mysterious Fermi-LAT GeV $\gamma$-ray excess is due to an unresolved population of millisecond pulsars (MSP) in the Galactic bulge, one expects this very same population to shine in X rays. For the first time, we address the question of what is the sensitivity of current X-ray telescopes to an MSP population in the Galactic bulge. To this end, we create a synthetic population of Galactic MSPs, building on an empirical connection between $\gamma$- and X-ray MSP emission based on observed source properties. We compare our model with compact sources in the latest \Chandra\ source catalog, applying selections based on spectral observables and optical astrometry with \Gaia. We find a significant number of \Chandra~sources in the region of interest to be consistent with being bulge MSPs that are as yet unidentified. This motivates dedicated multi-wavelength searches for bulge MSPs: Some promising directions are briefly discussed.
\end{abstract}

\maketitle

\section{Introduction}
\label{sec:intro}

A mysterious excess, discovered at GeV energies in the data 
of the Large Area Telescope (LAT) onboard the \Fermi~satellite, has been thrilling scientists for more than a decade.
The so-called \Fermi~GeV excess has been thoroughly characterized by several, independent, groups, see e.g.~\cite{Abazajian:2012pn,Gordon:2013vta,Calore:2014xka,Daylan:2014rsa,TheFermi-LAT:2015kwa}. Its spectral energy distribution is peaked at about 2 GeV, resembling the cumulative emission of known millisecond pulsars (MSPs)~\cite{Abazajian:2010zy} or what is expected from dark matter particles annihilating into 
high-energy photons, see e.g.~\cite{Calore:2014nla}. 
The GeV excess spatial distribution, which may entail stronger implications for its nature, is instead more debated.
It was initially found to match what is predicted by dark matter annihilation models~\cite{Daylan:2014rsa,Calore:2014xka}. However, more advanced and technically refined analyses showed that the $\gamma$-ray excess emission more closely traces old stars in the Galactic bulge~\cite{Bartels_bulgelum, Macias:2016nev,Macias:2019omb}.
This finding
supports the possibility that the excess is caused by a large population of MSP-like $\gamma$-ray
emitters in the Galactic bulge, too faint to be detected
as individual sources by the LAT (i.e.~{\it unresolved}).
Nonetheless, some doubts are still cast on the excess morphology~\cite{DiMauro:2021raz}.
Analyses of 
photon counts statistics may potentially shed light 
on the nature of GeV excess $\gamma$ rays, by discriminating 
point source and diffuse emission contributions. 
While early works~\cite{Bartels:2015aea,Lee:2015fea} appeared to corroborate a point-source nature of the GeV excess, the dark matter interpretation was revamped~\cite{Leane:2019xiy,2020arXiv200212370L}, because of yet unexplored systematics affecting photon counts statistical methods~\cite{Chang:2019ars,Buschmann:2020adf,Zhong:2019ycb}.
In spite of that, a recent work showed evidence for the presence of unresolved point sources partially contributing to the excess, and a preference for a bulge-like morphology~\cite{Calore:2021bty}. 
Neural networks techniques have also been shown to be promising in identifying sub-threshold $\gamma$-ray point sources~\cite{Caron:2017udl,List:2020mzd}. 

To conclusively prove the nature of the \Fermi~GeV
excess, a multi-wavelength approach can allow us to test (and constrain) the ``unresolved MSPs'' hypothesis.
Predictions for radio observations with current and future telescopes~\cite{Calore_radioprospects_msp}
have contributed to propel an on-going observational effort with radio interferometers, such as the Very Large Array (VLA) and MeerKAT, to look for radio counterparts of the \Fermi~GeV excess. Future multi-messenger probes involving gravitational waves have also been discussed~\cite{Calore:2018sbp}.
For sure, MSPs also emit X rays through thermal (from heating 
of magnetic polar caps) 
or non-thermal (e.g.~from relativistic particle acceleration in the pulsar magnetosphere, or shock-driven interactions between pulsar wind and companion material in binaries)  mechanisms~\cite{2018IAUS..337..116B}.
Several X-ray analyses have
targeted known radio and/or $\gamma$-ray MSPs to look 
for X-ray counterparts, e.g.~\cite{2011ApJ...733...82M,Marelli:2015vsa, Lee_Lx}.
The most complete census of known X-ray MSPs~\cite{Lee_Lx} spectrally characterized about 50 MSPs with data from \Chandra, {\it XMM-Newton}, {\it Suzaku}, {\it Swift}, {\it ROSAT}, and {\it BeppoSAX}. 

Building on the multi-wavelength emission of MSPs, 
here we assess {\it for the first time}
what is the sensitivity of current X-ray telescopes, notably \Chandra, to a Galactic bulge MSP population which would be responsible for the \Fermi~GeV excess.
While $\gamma$-ray data are not sensitive yet
to the detection of individual bulge MSPs~\cite{Bartels_mspluminosity,Bartels:2017xba}, {\it can the available deep X-ray observations of the inner Galaxy unveil them?}
To answer this question, we create a synthetic population of Galactic MSPs, which includes 
contribution from an MSP bulge component modeled such as to match spatial and spectral properties of the GeV excess from~\cite{Bartels_bulgelum}, Sec.~\ref{sec:msppop}. 
X-ray predictions are inferred via
an empirical connection between $\gamma$- and X-ray MSP emission based on~\cite{Lee_Lx}, Sec.~\ref{sec:msppop}. 
In Sec.~\ref{sec:chandra}, we present the \Chandra~source catalog and the corresponding sensitivity map.
In Sec.~\ref{sec:det_MSP}, we use the latter to define and characterize the detectable MSP population.
In Sec.~\ref{sec:det_Chandra}, we discuss X-ray spectral cuts as well as the complementary information provided by \Gaia\ astrometric observations.
In Sec.~\ref{app:MC_sys} and~\ref{app:cat_sys}, we investigate the systematic uncertainties associated with our model and with the \Chandra~catalog we use. Finally, in Sec.~\ref{sec:concl} we present future prospects and conclude.
Our ultimate goal is to understand how far we are from a possible discovery and, anticipating our encouraging findings, to promote dedicated multi-wavelength searches for bulge MSPs.

\section{The Galactic MSP population}
\label{sec:msppop}
We consider the Galactic MSPs to be composed by an observationally rather well-constrained disk component, 
 plus the elusive population in the Galactic bulge, putative origin of the \Fermi~GeV excess.

\subsection{The $\gamma$-ray population modeling}
\begin{figure*}[ht!]
\includegraphics[width =1.\textwidth]{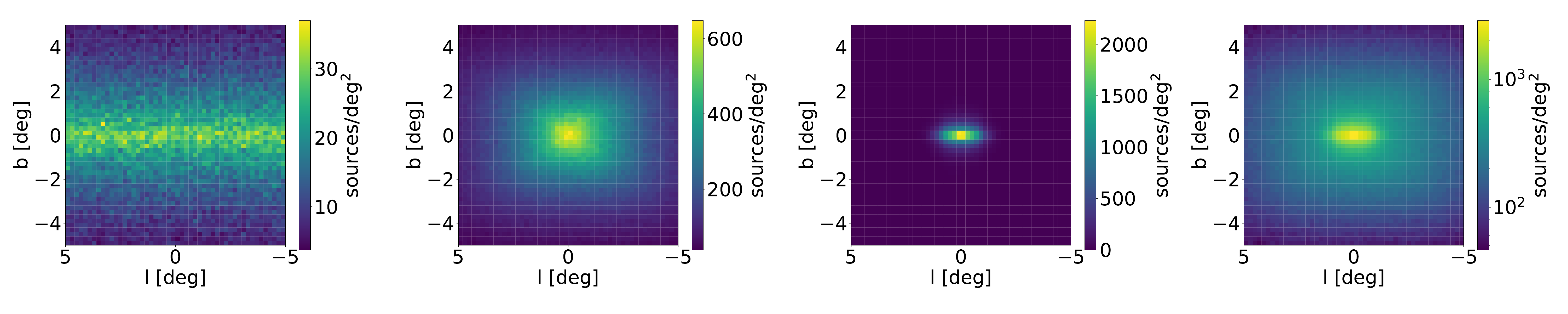}
\caption{From left to right, source density of the MSP Galactic population for the disk, BB, and NB components, and their sum.
}  
\label{fig:density_source}
\end{figure*}

We base the modeling of disk MSPs on a recent analysis of 96 \Fermi-LAT identified $\gamma$-ray MSPs~\cite{Bartels_mspluminosity}.
For the MSP disk spatial distribution, we adopt the ``Lorimer-disk'' best-fit profile, see Eq.~\ref{disk_number_density} in Appendix~\ref{app:MSP_pop}. 
The best-fit model for the ($0.1-100$ GeV) $\gamma$-ray luminosity function (GLF) of disk MSPs was found there to be a broken power law, see Eq.~\ref{eq:GLF}.
From the estimated best-fit average disk luminosity and using the best-fit broken power-law GLF,
we can compute the average total number of disk MSPs, see Tab.~\ref{tab:number_MSP}.
We show the spatial distribution of MSP disk source density in the leftmost panel of Fig.~\ref{fig:density_source}.

The modeling of bulge MSPs is inspired by observations of the GeV excess.
In particular, their spatial distribution builds upon the 
results of~\cite{Bartels_bulgelum}, and follows the morphology of red clump giants in the boxy bulge (BB)~\cite{Cao_bar_models} and of infrared observations of the nuclear bulge (NB)~\cite{Launhardt02}. The NB is, in turn, composed by the nuclear stellar disk (NSD), and nuclear stellar cluster (NSC) (equations provided in Appendix~\ref{app:MSP_pop}).
We show the source density spatial distribution of the different bulge components in Fig.~\ref{fig:density_source}.
The BB extends approximately from $30\degree$ to $-20\degree$ in longitude and from $-20\degree$ to $20\degree$ in latitude, well beyond the boundaries of Fig.~\ref{fig:density_source}.
From that figure, we can also see that 
the NSD is instead located between $|l|<2\degree$ and $|b|<2\degree$, and the 
NSC is contained in the innermost $2\degree \times 2 \degree$. 

As we will comment below, given the very narrow region of interest (ROI) considered for this study, the uncertainties related to the choice of the GeV excess morphology only have a minor impact on the final results.
Although it is difficult to constrain the GLF of bulge MSPs
due to the lack of resolved $\gamma$-ray objects,
existing studies have found that it is consistent with the GLF of resolved disk MSPs~\cite{Ploeg:2017vai, 2020arXiv200810821P}.
We therefore assume the GLF of 
bulge MSPs to be the same as the one for the disk population. We will show that our predictions are only mildly affected by changes
of the GLF parameters, see Sec.~\ref{app:MC_sys}.
Fixing the GLF and imposing 
that the total average luminosity of the Galactic bulge component 
matches the best-fit estimates from~\cite{Bartels_bulgelum}, 
the total number of sources in the BB and NB are found to be
27674 and 2700, respectively.
 
We report these numbers in Tab.~\ref{tab:number_MSP}.

From GLF and source spatial distribution, we can simulate a corresponding $\gamma$-ray energy flux
for each synthetic source in the $0.1-100$ GeV band.
\begin{table}[!ht]
    \centering
    \caption{Observed or estimated average $\gamma$-ray luminosity $\langle L_\gamma^{\rm obs} \rangle$~\cite{Bartels_mspluminosity, Bartels_bulgelum} for the Galactic MSP population components, together with the derived total number of MSPs in each component, $N_{\rm tot}$.}
        \begin{tabular}{c|c|c}
        \hline\hline
     & $\langle L_\gamma^{\rm obs} \rangle$ [erg/s] & $N_{\rm tot}$\\[2pt]\hline
    BB & $1.73\times10^{37}$ & 27674 \\
    NSD & $1.63\times10^{36}$ & 2606 \\
    NSC & $5.89\times10^{34}$ & 94 \\
    Disk & $1.5\times10^{37}$ & 24009 \\
    \hline\hline
\end{tabular}
\label{tab:number_MSP}
\end{table}

\subsection{The X-ray flux distribution}
Our leading working hypothesis is that we can predict the X-ray MSP emission from the 
$\gamma$-ray one.
Many studies exist on the correlation between X-ray and $\gamma$-ray
luminosities and pulsars' spin-down power, see e.g~\cite{Possenti_Lx_Edot, Kalapotharakos_Lg,2019PASP..131b4201W}.
One viable strategy would be to build on this double correlation, and model the $\gamma$ and X-ray emission through the pulsars' spin-down power.
Given the large uncertainties present in each correlation, we prefer not to rely on those, but rather to follow a more observation-driven approach. 
Similarly, we avoid to rely on multi-wavelength emission models of the MSPs spectral energy distribution, given the limited sample over which such multi-wavelength studies have been performed, and the variety of models that can be fitted to pulsars' spectra, see~\cite{2020MNRAS.492.1025C,2019MNRAS.489.5494T}.
In order to predict X-ray fluxes of our synthetic sources, we therefore rely on an empirical connection between \textit{observed} $\gamma$- and X-ray MSP emission properties.
We notice that all the above-mentioned approaches, including the one followed here, rely on the assumption that the {\it observed} sample of sources is representative of the underlying population, at least over the energy flux range supported by data. Deriving the properties of the unresolved population starting from observed sources is common practice in $\gamma$-ray astrophysics, see, as an example, the predictions of the contribution of blazars and star-forming galaxies to the \Fermi-LAT $\gamma$-ray diffuse background~\cite{Ajello:2015mfa, DiMauro:2013zfa, Ajello:2020zna}.

Ref.~\cite{Lee_Lx} is, to our knowledge, the most complete census of X-ray MSPs, and presents power-law spectral fits to 47 detected X-ray MSPs.
By cross-correlating this sample with the latest release of the 4FGL catalog~\cite{4FGL}, we found that 40 objects have $\gamma$-ray \Fermi-LAT counterparts, 
possibly implying that not all observed X-ray MSPs have a $\gamma$-ray detected counterpart.
The number of MSPs emitting in $\gamma$ rays and the one emitting in X rays may differ, for instance due to the different (and poorly constrained) emission geometries~\cite{Marelli:2015vsa}. 
However, based on our calibration sample, we conservatively assume 
that each $\gamma$-ray MSP in our Monte Carlo simulation also has associated X-ray emission. By doing so, we are not over-predicting the number of possible X-ray detections.

In order to predict X-ray fluxes of interest for this study, we identify two variables which are relevant for making this prediction.

\begin{figure}
\includegraphics[width=0.45\textwidth]{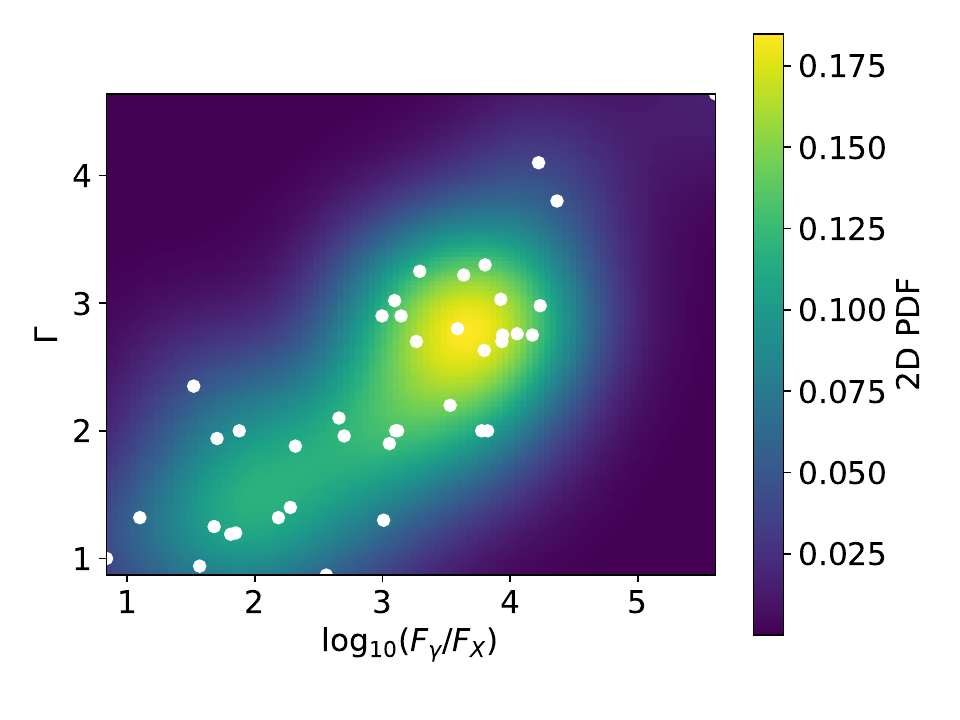}
\caption{KDE 2D joint PDF (colored background) of $\log_{10}(F_\gamma/F_X)$ and X-ray spectral index $\Gamma$ from the 40 X-ray observed MSPs having a $\gamma$-ray counterpart. Original data from~\cite{Lee_Lx} are shown by the white dots.}  
\label{fig:gammaX_kde}
\end{figure}

First, we consider the $\gamma$-to-X flux ratio, $F_\gamma/F_X$, of the 40 X-ray MSPs having $\gamma$-ray counterpart, where $F_\gamma$ is the $0.1-100$ GeV energy flux and $F_X$ the $2-10$ keV {\it unabsorbed} 
energy flux. 
Modeling such a ratio and knowing $F_\gamma$ for each synthetic source, we can then generate a corresponding X-ray flux.
The second variable of interest is the X-ray spectral index, $\Gamma$, provided by~\cite{Lee_Lx}. Extracting X-ray spectral indices for our simulated sources would allow us to model $F_X$ in any energy band, in particular the ones covered by \Chandra. We notice that
$\Gamma$ is significantly correlated with $\log_{10}(F_\gamma/F_X)$ (Spearman coefficient of 0.782). We therefore build a 2D probability distribution function (PDF) of $\log_{10}(F_\gamma/F_X)$ and $\Gamma$ from the 40 MSPs. Given the paucity of data, we use a kernel density estimation (KDE) algorithm~\cite{scikit-learn} to derive the joint PDF, checking the stability of the result against the bandwidth choice and the optimization algorithm.
From this PDF, displayed in Fig.~\ref{fig:gammaX_kde},
we extract an
$F_\gamma/F_X$ ratio and index $\Gamma$ for each synthetic source, and, from there, X-ray fluxes in any \Chandra~energy band.
Some uncertainties on the modeling of the $\gamma$-X correlation will be tested in Sec.~\ref{app:MC_sys}.
We also stress that, in what follows, we will not extrapolate our model beyond the validity range of the $\gamma$-X correlation, as directly supported by the data points.

The {\it absorbed} differential photon flux per unit of energy is 
obtained by modeling the Galactic absorption:
\begin{equation}
\begin{split}
    S^{\rm abs}(E) &= S^{\rm unabs}(E) \times \exp{(- N_H \sigma(E))}\\
    &=  A \, (E / 1 \, \rm keV)^{-\Gamma} \times \exp{(- N_H \sigma(E))} \, , 
\end{split}
\end{equation}
where $S^{\rm unabs}(E)$ is the unabsorbed photon flux, 
modelled by a power law with amplitude $A$ and spectral index $\Gamma$. 
$N_H$ (cm$^{-2}$) is the total hydrogen column density along the line of sight, 
and $\sigma (E)$ the photoelectric absorption cross section. 
We parameterize $\sigma(E)$ as in~\cite{BCMC92} with Galactic elemental abundances
 from~\cite{angr89}. 
To build the hydrogen column density $N_H$, we use the publicly available gas maps adopted in~\cite{Macias:2016nev}, and publicly available at \url{https://github.com/chrisgordon1/galactic_bulge}.
These maps are obtained from atomic (HI) and molecular (H2) hydrogen surveys~\cite{LABsurvey,Dame:2000sp} with an hydrodynamic approach, which accounts for non-circular gas motion in the inner Galaxy.
Since hydrodynamic maps provides better kinematic
resolution towards the inner Galaxy than standard deconvolution methods~\cite{Pohl:2007dz}, we adopt those as baseline hydrogen model. 
They are split in four concentric rings, separated by R = 3.5, 8 and 10 kpc, providing a coarse-grained 3D hydrogen distribution. 
We take the hydrogen-to-CO conversion factor partially from~\cite{Macias:2016nev}, using $X_{CO} = 0.4 \, (1.0) \times 10^{20}$ cm$^{-2}$/(K km s$^{-1}$) for R $\leq$ 3.5 kpc ($3.5\, \rm kpc <$ R $\leq$ 8.0 kpc).
The outer rings (R $>$ 8 kpc) X$_{CO}$ being poorly or completely un-constrained by~\cite{Macias:2016nev}, 
we adopt the standard reference value of $1.9 \times 10^{20}$ cm$^{-2}$/(K km s$^{-1}$)~\cite{2012ApJ...750....3A}.
The total column density being $N_H = N_{HI} + 2N_{H2} + N_{\rm dust}$, we also include the contribution 
from the dark neutral medium~\cite{2005Sci...307.1292G} by including the dust-to-gas residual reddening maps 
from~\cite{Macias:2016nev}.
To convert E(B-V) residual maps in units of hydrogen column density we use a 
dust-to-gas ratio $X_{\rm dust}= 41.4 \times 10^{20}$ cm$^{-2}$mag$^{-1}$~\cite{Casandjian15}.

We test other choices for the modeling of the total hydrogen column density in Sec.~\ref{app:MC_sys}.

\section{\Chandra~source catalog and sensitivity map}
\label{sec:chandra}
With its unique high spatial resolution and low
instrumental background, \Chandra~is an excellent instrument
to image the X-ray sky and detect X-ray sources in the 
$0.1-10$ keV energy band~\cite{Weisskopf:2001uu}. 
\Chandra~is equipped with two imaging detectors: The 
Advanced CCD Imaging Spectrometer (ACIS), and the High Resolution Camera (HRC).

For the purpose of this work, we use the latest 
release of the \Chandra~Source Catalog (CSC 2.0, CSC hereinafter)
of X-ray sources~\cite{2010ApJS..189...37E,2018AAS...23123802P}.
The catalog provides observed properties in multiple energy bands for 
about 320000 compact and extended X-ray sources, 
as well as details of stacked-observation and detection regions.

\begin{figure}[!thb]
\includegraphics[width=0.48\textwidth]{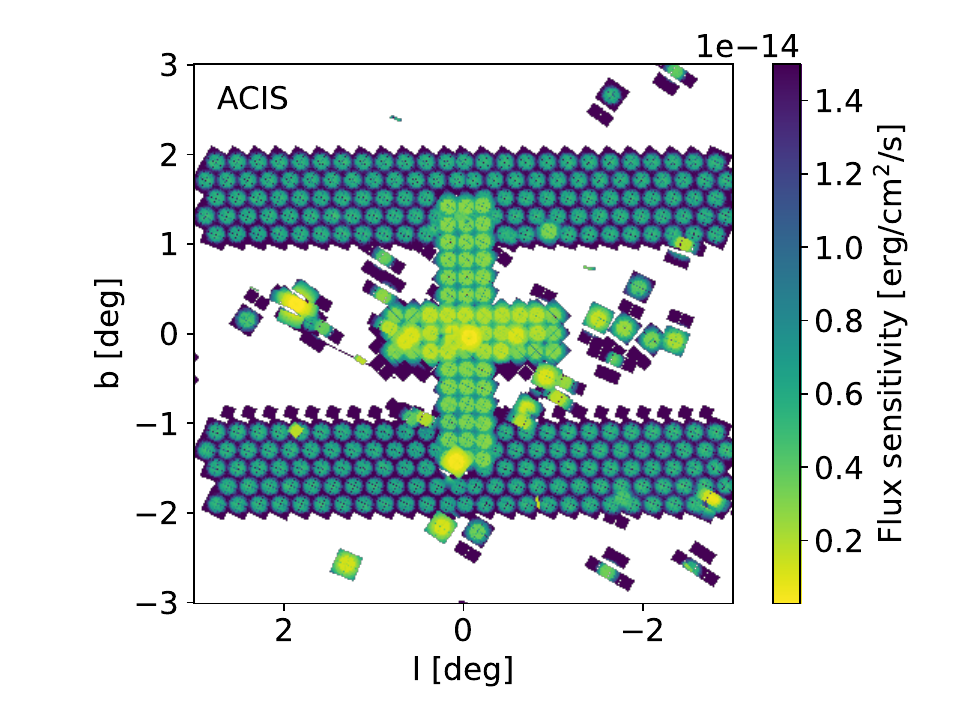}
\caption{ACIS limiting sensitivity map for the \texttt{TRUE} and \texttt{MARGINAL} detection likelihood classes. White regions are not covered by the \Chandra~observations used to build the CSC catalog.
}
\label{fig:sens_maps}
\end{figure}

Among the CSC data products, multi-band limiting sensitivity maps are available.
We focus on an ROI of 6$\degree \times 6 \degree$ about the Galactic center, and retrieve 
 sensitivity maps from the \Chandra~data base \url{https://cxc.harvard.edu/csc/columns/limsens.html}, binned with a 1$\times$1 arcmin$^2$ pixel size in $(l,b)$. 
Our baseline sensitivity map, displayed in Fig.~\ref{fig:sens_maps}, corresponds to the estimated minimum energy flux in the ACIS broad band ($B$, $0.5-7.0$ keV) for a source to be detected and classified as \texttt{TRUE} or \texttt{MARGINAL} at the detection position, where 
the source detection likelihood classes are defined at \url{https://cxc.harvard.edu/csc/columns/stack.html}.
 All predictions which follow, for both Monte Carlo and \Chandra\ catalog, refer to this specific ACIS sensitivity map.
We note that 
 CSC source detection
 is not based on likelihoods derived from Poisson fluctuations, like
 those used to build the sensitivity maps. Therefore, for a meaningful comparison between Monte Carlo and catalog, we apply the sensitivity cut also to CSC sources, see ``Limiting Sensitivity'' at~\url{https://cxc.harvard.edu/csc/char.html}, and~\url{https://cxc.harvard.edu/csc/memos/files/Primini\_limiting-sensitivity.pdf}.
The choice of the 
detection likelihood class only mildly impacts the selection of CSC sources, see Sec.~\ref{app:cat_sys}.

\section{Characterization of detectable MSPs}
\label{sec:det_MSP}
\begin{figure}
\includegraphics[width=0.48\textwidth]{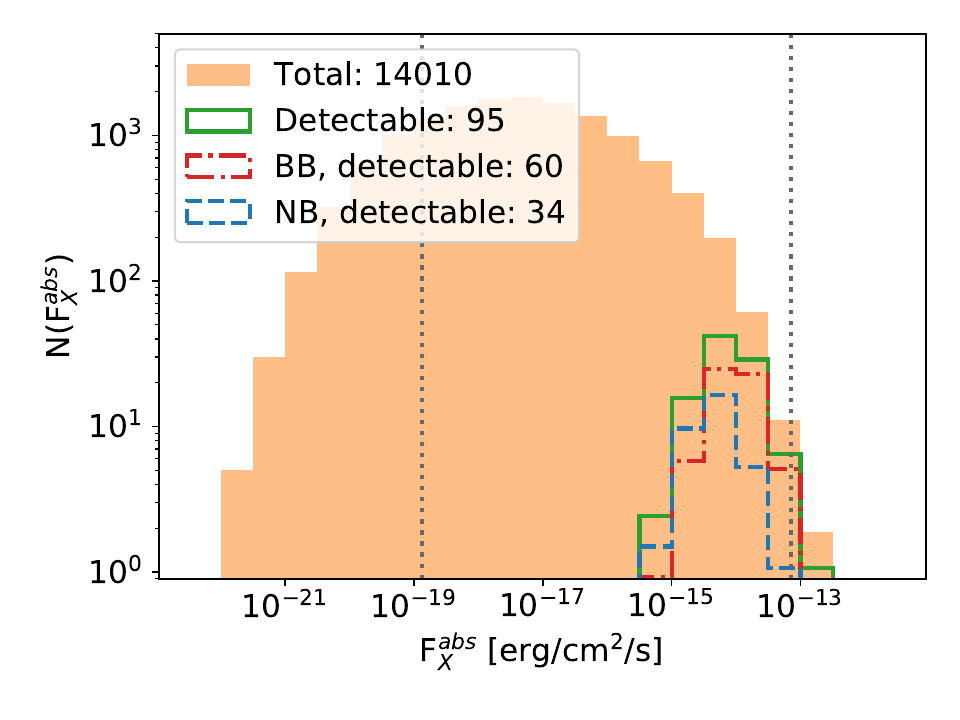}
\caption{X-ray energy flux distribution ($0.5-7$ keV) of the synthetic MSP population, averaged over 100 Monte Carlo simulations: Total MSPs in the ROI (orange filled), total detectable MSPs (green solid) including MSPs from BB (red dot-dashed), NB (blue dashed) and disk (not shown). The vertical dotted lines illustrate the validity range of our model extrapolation (see text for details).
Errors from Monte Carlo dispersion are not shown here for clarity, see text for details.
}
\label{fig:hist_det}
\end{figure}

\begin{figure}
\includegraphics[width=0.48\textwidth]{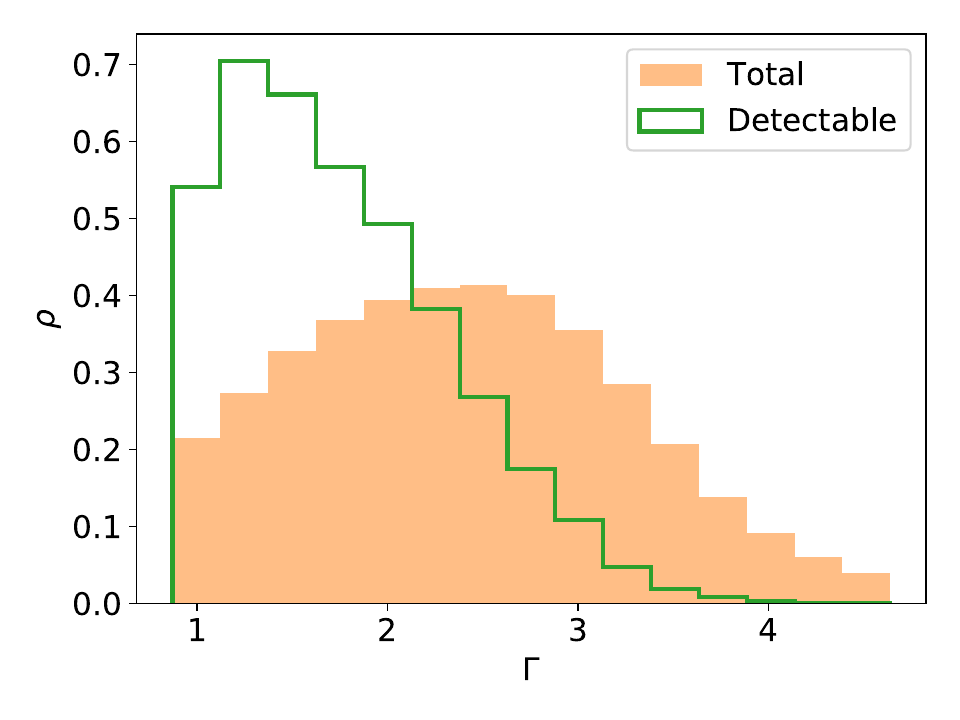}
\caption{Spectral index density histograms for all MSPs in the ROI (oranged filled) and all detectable MSPs (green solid).
}
\label{fig:hist_gam}
\end{figure}

Averaging over 100 Monte Carlo simulations of the Galactic MSP population, we find a total of 14010 $\pm$ 91\footnote{The estimated errors come from the dispersion over 100 Monte Carlo simulations, unless stated otherwise.} MSPs in the chosen ROI, as displayed by the orange histogram in Fig.~\ref{fig:hist_det}.
We then compute the number of ``detectable'' MSPs that have absorbed energy flux larger than the \Chandra~sensitivity at the source position. 
We obtain 60 $\pm$ 7 MSPs detectable from the BB, 34 $\pm$ 6 from the NB, and 1$\pm$1 from the disk, adding up to a total of 95 $\pm$ 9 detectable MSPs.
The contribution to the detectable MSPs from the disk component is negligible in our ROI. 
We indeed predict a total of 700 $\pm$ 27 disk MSPs in the ROI, 
of which, on average, only 1 is detectable. 
In our ROI, more than 90\% of the disk population is located behind the Galactic center, on the other side of the Galaxy, and so hardly detectable because of Galactic absorption and distance. Moreover, the disk is denser between -1$^\circ$ and 1$^\circ$ in latitude, and this band is less covered by \Chandra~observations than regions with $|b| > 1^\circ$.
The energy flux distributions of the total Galactic
MSPs population and its bulge detectable components are
shown in Fig.~\ref{fig:hist_det}.
In particular, the green histogram displays the total number of detectable MSPs; the rounded entries (Monte-Carlo dispersion errors) associated with its 6 bins ranging from $10^{-15.5}$ to $10^{-12.5}$ erg/cm$^2$/s are: 2 (1), 15 (3), 42 (6), 28 (5), 6 (2), 1 (1).
Between vertical dotted lines in Fig.~\ref{fig:hist_det}, we also highlight the most credible interval of our model, where the $\gamma$-to-X correlation is directly supported by data: The left and right dotted lines are the fluxes that correspond to the minimal and maximal luminosity of observed X-ray MSPs, respectively, if we were to project those sources at the Galactic center.
The detectable sources naturally tend to be the brighter ones, as can be seen in Fig.~\ref{fig:hist_det}, but also the harder ones. In the ACIS broad band, sources with lower spectral indices are less affected by absorption and therefore have a larger flux than sources with higher spectral indices. Moreover, the observed correlation between the flux ratio $\log_{10}(F_\gamma/F_X)$ and the spectral index $\Gamma$ favors high $F_X$ for low $\Gamma$ (see Fig.~\ref{fig:gammaX_kde}). The mean MSP spectral index in our ROI is 2.41, while it drops to 1.76 in the detectable population. The spectral index distribution for the two populations is shown in Fig.~\ref{fig:hist_gam}. 
The mean distance to detectable MSPs, 8.48 kpc, is slightly smaller than the mean distance to ROI MSPs, 8.85 kpc. The latter is larger than the distance between the Sun and the Galactic center (8.5 kpc) because of the volume of the ROI: More sources behind the Galactic center than in front of it falls in the ROI considered. Finally, the column density distribution shows a clear dichotomy between the BB MSPs, with a mean value of $2.92\times10^{22}\rm\, erg/cm^{2}/s$ and the NB MSPs, with a mean value of $6.81\times10^{22}\rm\, erg/cm^{2}/s$. In Sec.~\ref{app:MC_sys}, we comment about the robustness of the characteristics of the detectable population against systematic uncertainties associated with the model.

\section{\Chandra~candidates selection}
\label{sec:det_Chandra}
For a meaningful comparison between Monte Carlo and \Chandra~catalogs, from the CSC
we select non-variable compact sources whose energy 
flux in the ACIS wide band, i.e.~\texttt{flux$\_$aper90$\_$b}, 
 is larger than the limiting sensitivity at
 the source position.
With these minimal cuts we select 6918 sources in our ROI, including 6837 sources having at least one intra-wide-band flux information provided. Hence, according to our model, detectable MSP sources represent 1.4\% of the full \Chandra\ catalog in the ROI of interest. We show below that this fraction can be significantly enhanced with appropriate spectral and distance cuts.

\subsection{Spectral constraints with \textit{Chandra}}
\label{sec:spec_obsvb}
In order to exploit the X-ray spectral information and reject \textit{Chandra} candidates unsuitable to be MSPs, we define the {\it flux ratios}:
\begin{equation}
    \phi_{ij} = \frac{F_i-F_j}{F_i+F_j} \, ,
\end{equation}
where $F_i$ is the absorbed energy flux in the $i$ band (\texttt{flux$\_$aper90$\_$i} in the CSC): Hard ($H$, $2-7$ keV), medium ($M$, $1.2-2$ keV), and soft ($S$, $0.5-1.2$ keV). 
We also introduce the {\it band fractions}:
\begin{equation}
    \beta_i = \frac{F_i}{F_B} \, ,
\end{equation}
where $i$ refers to the $H$, $M$ or $S$ bands defined above, and $B$ is the ACIS broad band. 
From the simulated (absorbed) energy fluxes, we calculate these quantities for the detectable bulge MSP population. 
From over 100 Monte Carlo simulations, the extreme ranges of MSP spectral observables are: $-0.066 < \phi_{HM} < 1$, $-0.015 < \phi_{HS} < 1.$ and $0.051 < \phi_{MS} < 1$, and
$0.32<\beta_H<1$, $0.00015<\beta_M<0.44$ and $0<\beta_S<0.33$.

\subsection{Optical astrometry with \Gaia} 
The \Gaia\ ESA mission~\cite{2016A&A...595A...1G} provides $\mu$-arcsec astrometry for more than 1 billion stars down to magnitudes of about 20 in the white-light
G band (330–1050 nm), complemented by 
radial-velocity and photometric information.
The latest \Gaia\ data release DR3~\cite{gaia_edr3} contains positions and G band magnitudes for 
1.8 billion sources. Among them, about
1.5 billion sources possess parallaxes. 
Distances have been determined, when possible, using a probabilistic approach, and a self-consistent, reduced, catalog has been compiled~\cite{Bailer-Jones_eDR3}. The latter contains geometric distances, which are the most numerous, and photo-geometric distances, which are less numerous but more precise.
In what follows, we make use of this \Gaia\ catalog, and the photo-geometric distances if not stated otherwise. 

A recent study of optical counterparts of pulsars in the ATNF catalogue~\cite{Antoniadis:2020gos} has revealed that only 18 MSPs, in binary systems, out of 107 MSP show an optical \Gaia\ counterpart.
This sample is mostly local, indicating that MSP companions are typically rather dim with apparent magnitude between $18-20$ at distances of $1-2$ kpc. 
While MSPs in the Galactic disk may therefore possess 
optical \Gaia\ counterparts, MSPs in the bulge, at $4-10$ times higher distances, should be invisible for \Gaia.
We use the presence of optical counterparts and distance information 
to further reduce our sample of sought-after {\it bulge} MSP candidates, 
knowing that the distance between the Sun and the detectable bulge MSPs covers 5.24 kpc $< d <$ 11.98 kpc as extracted from our Monte Carlo simulations. We define a positive cross-match whenever a \Gaia\ source is found within the 95\% C.L. semi-major axis of the error ellipse of a CSC source, \texttt{err$\_$ellipse$\_$r0}.
Out of the 6918 sources of our initial catalog, we find 2093 \Chandra-\Gaia\ positive cross-matches. Using the geometric distance when the photo-geometric distance is not available, we find 131 additional matches, so 2224 positive cross-matches in total. 

\subsection{Conservative and aggressive selections}
\label{sec:csc_selection}
We make use of \textit{Chandra} spectral constraints and \Gaia~astrometry to further reduce the sample of
CSC sources of interest.
We define two different selections.

\smallskip

\paragraph{{\bf Conservative selection.}}
This selection of CSC sources is meant to reject most of the \Chandra~sources that 
we can safely say are {\it not compatible} with spectral and distance distributions of detectable bulge MSPs.
As such, the conservative selection allows us to assess if the MSP population model is excluded or not by the \Chandra~catalog.
To the 6918 non-variable, non-extended sources above the sensitivity threshold in our \Chandra\ ROI, we impose that:
i) Whenever a spectral observable is available, the source is retained only if its value falls within the corresponding Monte Carlo-deduced range. If intra-wide-band fluxes are zero or unavailable (as typically occurs for too dim sources), the source is {\it kept}. This reduces the sample to 3606 objects.
ii) If the distances of all \Gaia\ counterparts are either closer than the distance to the bulge or further away, i.e.~the source is in the disk, the source is {\it rejected}.  
This further reduces the selected sources down to 3153. For illustration, by reducing the cross-matching radius to 1 arcsec (of the order of the systematic error on the positional reconstruction), we get 1289 cross-matches between the 6918 \Chandra\ and \Gaia~catalogs, ending up with 3260 sources in our conservative selection.

\smallskip

\paragraph{{\bf Aggressive selection.}}
This second selection aims at isolating the {\it most promising} sample of bulge ``MSP-like'' candidates.
To this end, we keep sources:
i) For which all $\phi_{ij}$'s and $\beta_i$'s are computable and fall within our Monte Carlo extreme intervals. This reduces the sample to 589 objects.
ii) That have {\it no cross-matches} with \Gaia~sources, following the rationale discussed above. By doing so, we discard sources that are surely in the disk (193), sources that may be in the disk or in the bulge (26), and, finally, also sources that are surely in the bulge (85). This reduces the sample to only 285 objects. Such a selection is not based on \Gaia\ distance information, and therefore the full \Gaia\ DR3 catalog~\cite{gaia_edr3} can also be used. 
In this case, the aggressive sample would reduce to 203 candidates.

\bigskip

\begin{figure}
\includegraphics[width=0.48\textwidth]{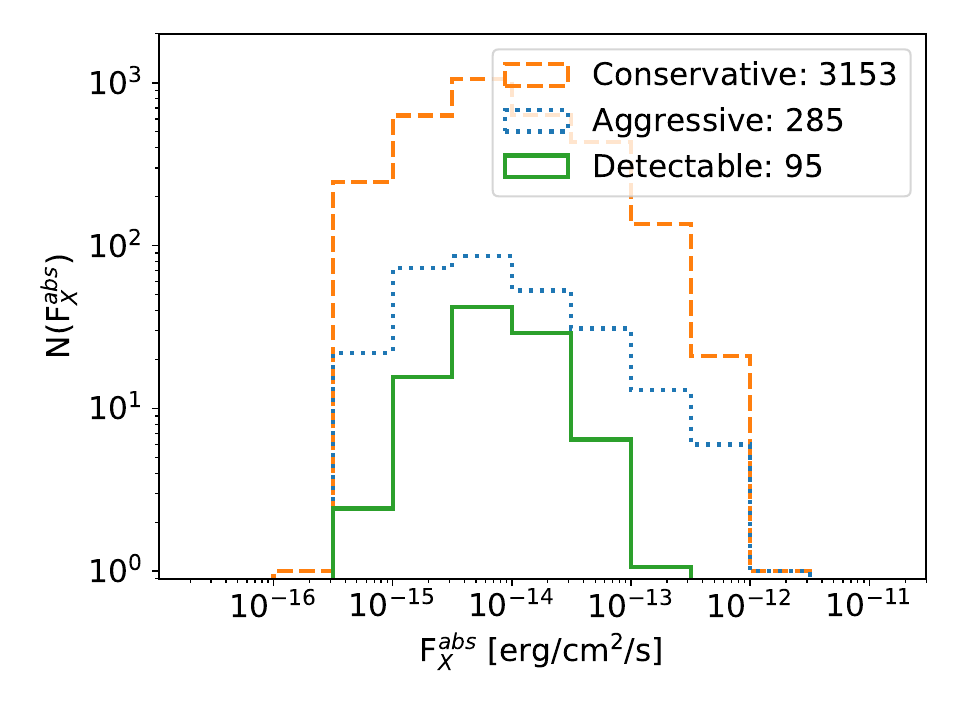} \\ 
\caption{As in Fig.~\ref{fig:hist_det}, showing Monte Carlo predictions for the total number of detectable MSPs (green solid), together with the conservative (orange dashed) and the aggressive (blue dotted) CSC selections.}
\label{fig:hist_det_2}
\end{figure}

We display the energy flux distribution of our two selected samples and of our detectable Monte Carlo sample in Fig.~\ref{fig:hist_det_2}.
Such a figure has the illustrative scope of showing that, by
comparing the flux distributions of our synthetic MSP population and of CSC selected candidates, this {\it simple} bulge MSP model is not (yet) excluded by the X-ray conservative selection. Moreover, the aggressive approach allows to reveal a fairly limited sample of promising targets suitable for further investigation, as discussed in the next section.  
Further improvements in the selection of
MSP-like candidates can be achieved, for example, by
cuts on the flux interval.
These conclusions hold true against several systematic uncertainties related either to the model (Sec.~\ref{app:MC_sys}) or to the selection itself (Sec.~\ref{app:cat_sys}).
%

\section{Systematic uncertainties associated with the model}
\label{app:MC_sys}
In this section, we explore the main uncertainties that can alter the number of detectable synthetic MSPs, and show that our predictions are robust against the systematics highlighted in what follows.
In particular, we investigate variations of bulge GLF parameters, of the parameterization of the $\gamma$-X correlation, as well as of the hydrogen column density maps.
We find that our main conclusions hold true against these systematics, namely we still have (i) a significant number of detectable MSPs, and (ii) a promising sample of targets provided by our aggressive selection.
We summarize these tests in Tab.~\ref{tab:sys_summary}.

We notice that, although there is still some uncertainty in the morphology of the GeV excess, modeling the central MSP population with, for example, a spherically symmetric dark matter-inspired distribution instead of the bulge one will not change significantly the total number of detectable sources given the very narrow region around the Galactic center we focus on, where the two distributions are very much compatible. We therefore do not include the variation of the GeV excess morphology among our tests.

\subsection{$\gamma$-ray luminosity function for the bulge}
Since current $\gamma$-ray data are not sensitive
to the detection of individual MSPs in the Galactic bulge, 
it is difficult to robustly constrain 
the GLF of bulge MSPs.
Although the GLF of the putative MSPs in the Galactic bulge
has been found to be consistent with that characterizing resolved disk MSPs~\cite{Ploeg:2017vai}, we cannot exclude
that the GLF of bulge MSPs differs from the disk one.
We here test this possibility and the impact that a variation of the bulge GLF can have on X-ray sensitivity prospects. 
We vary the parameters of our baseline GLF around their best-fit values~\cite{Bartels_mspluminosity}, but beyond the statistical 1$\sigma$ errors, and we check {\it a posteriori} that the number of
detectable $\gamma$-ray bulge MSPs for that variation is still in agreement with findings from~\cite{Bartels_mspluminosity}, i.e.~a few detectable $\gamma$-ray bulge MSPs (adopting the \Fermi-LAT detection sensitivity model as in~\cite{Bartels_mspluminosity}). 
Thoroughly exploring $\gamma$-ray implications for more extreme variations of the bulge GLF is beyond the scope of the present work. 
By varying $\alpha_1$, $\alpha_2$, and $L_b$ (see Eq.~\ref{eq:GLF}) one at the time,
 we find that the number of detectable 
X-ray sources is mildly affected by changes of the
$\gamma$-ray modeling.

We consider, as exemplary variations of the bulge GLF parameters, the following cases: 
For $\alpha_1 = -0.97$, we get $101 \pm 10$ detectable sources; for $\alpha_2 = 2.4$, $84 \pm 9$ (averages performed over 100 Monte Carlo simulations); and for $L_b = 10^{32.8}$ erg/s, $78 \pm 9$ (average performed over 40 Monte Carlo simulations).
We checked that not only the total number of sources, but also their flux distribution is not altered by variations of the GLF parameters (as well as by the other systematic uncertainties we test below). In general, the distribution of detectable sources is peaked at about $10^{-14}$ erg/cm$^2$/s.
The reason why X-ray predictions are not too sensitive
to the bulge GLF
lies in the fact that the largest fraction of detectable X-ray MSPs has $\gamma$-ray fluxes close
to the best-fit $L_b$ value, and, therefore, we need very extreme variations of the parameters to 
induce a sizeable effect on the number 
of detectable X-ray MSPs. 
Such variations, however, are not allowed if we require consistency with $\gamma$-ray (non-)detection of individual bulge MSPs. 

Additionally, changes in the modeling of the synthetic population may also affect the Monte-Carlo-based cuts for flux ratios and band fractions, and, therefore, the final source selection. However, the characteristics of the detectable population are not significantly impacted by the variations of the GLF, and we find that under the tested variations of the GLF parameters the aggressive selection would change by no more than 15\%$-$16\%.

\subsection{$\gamma$-to-X correlation}
Besides the KDE approach, we consider a multi-variate normal distribution for the joint PDF of $\log_{10}(F_{\gamma}/F_X)$ and $\ln(\Gamma)$ as a way to test the stability of our predictions against the modeling of the $\gamma$-to-X correlation. The multi-variate PDF is not fitted to the data, rather we use the mean vector and the covariance matrix associated to the 40 data points as parameters of the 2D normal distribution. 
In this case, 
we find a total of $69\pm8$ detectable sources in the Monte Carlo simulation, including $41\pm7$ BB sources and $27\pm5$ NB sources. This represents less potential detections than with the KDE joint PDF but the candidate selection is also affected: Only 2912 sources remain in the conservative selection and 181 in the aggressive selection. 
In this case, the mean spectral index of the total population in the ROI is 2.26, slightly lower compared to our baseline model. 
Accordingly, also the mean spectral index of the detectable population slightly decreases compared to the baseline model, reaching 1.41. 
The other characteristics of the detectable population (N$_H$, distance and flux) are not significantly affected.

Conversely, assuming the two variables to be ``uncorrelated'' would instead increase the number of detectable sources, as well as the candidates selection. 
In this case, we obtain 122 $\pm$ 12 detectable sources from the Monte Carlo simulation, the aggressive selection is made up of 353 sources, and the conservative selection reaches 3338 sources. 
Unlike our baseline model, the uncorrelated test does not favor any spectral index for higher X-ray flux. The spectral index distributions in the ROI and the detectable population are then very similar in shape, with a mean value of 2.35 and 2.44 respectively. 
The baseline and the uncorrelated model both have about 90 detectable source with $\Gamma < 2.9$. In the baseline model, only sources with these lower spectral indices reach X-ray fluxes high enough to be detectable, while in the uncorrelated case high spectral indices can also be detected. This explains why there are more detectable MSPs in the uncorrelated case.
We stress however that the uncorrelated case is unrealistic and, as such, grossly overestimates both the number of detectable and candidate sources.

\subsection{X-ray Galactic absorption}
To explore the systematic uncertainty related to the modeling 
of the total hydrogen in the Galaxy and its
distribution along the line of sight, we consider two additional models for Galactic hydrogen
column density.
First, we use the \texttt{dustmaps} Python module\footnote{\url{https://dustmaps.readthedocs.io/en/latest/}.}, which implements the 2D extinction map from~\cite{SFD98}.
Being a 2D map, no distance information can be retrieved, and the absorption towards a given MSP may be overestimated. 
Using 2D dust map should provide a lower limit on the number of detectable MSPs.
Secondly, we consider a model with no absorption by setting $N_H = 0$ cm$^{-2}$.
In this extreme case, on the contrary, we grossly underestimate the absorption and get an upper limit on the number of detectable MSPs.
In particular, we obtain $75 \pm 8$ detectable MSPs for the 2D dust map, and $267 \pm 15$ for the unabsorbed case (averages over 100 Monte Carlo simulations). 
Uncertainties in the Galactic hydrogen modeling therefore do not diminish the final X-ray predictions by more than 20\%, while any enhancement is {\it theoretically} bound to be within a factor 2.8 of our benchmark. With no absorption, the mean spectral index of the detectable population is 2.09, much larger than 1.76 in our baseline model. This demonstrates the influence of the high column densities on the detectable population discussed in Sec.~\ref{sec:det_MSP}.

As for the aggressive selection, we get 368 source candidates when using the 2D dust map, and 41 candidates in the unrealistic case of no absorption.

\section{Systematic uncertainties associated with the observations}
\label{app:cat_sys}
In this section, we present some of the most important systematics associated with the \textit{Chandra} catalog, and how they affect the final \textit{aggressive} selection of X-ray sources.
The results are summarized in Tab.~\ref{tab:sys_summary}.

\subsection{\Chandra~sensitivity map}
We compute Monte Carlo predictions, minimal, and aggressive CSC selections
obtained by using the sensitivity map corresponding only to the \texttt{TRUE} detection likelihood class.
In this case, we get $54 \pm 7$ detectable MSPs (vs. $95\pm 9$ for the baseline scenario), and 4703 source in the minimal CSC selection (vs.~6918). 
If we consider the aggressive CSC selection, we get 234 candidates.
The variation induced by the use of a more restrictive sensitivity map is no more than a factor of two, affecting similarly the signal and the background.
Also in this case, the flux distribution of candidates and detectable sources is unaffected by the choice of the detection likelihood class.

\subsection{Source flux uncertainties}
The source flux provided in the \Chandra\ catalog are delivered with the 1-$\sigma$ upper and lower limits. However, our Monte Carlo simulation does not include uncertainties on the X-ray fluxes, so we decided to ignore them through our analysis. To demonstrate that this has no impact on our results, namely that the model is not excluded by the data and that we are able to select a reduced sample of promising candidates, we test the two following extreme cases.

Comparing the upper limit \texttt{flux$\_$aper90$\_$hilim$\_$b} instead of the flux \texttt{flux$\_$aper90$\_$b} to the limiting sensitivity, we obtain 4635 candidates in the conservative selection and 303 in the aggressive selection. We note that considering the upper limit instead of the flux itself leads to the inclusion of sources for which \texttt{flux$\_$aper90$\_$b} = 0. The band fraction of such sources cannot be computed, so they are kept in our conservative selection but excluded from our aggressive sample (see Sec.~\ref{sec:csc_selection}).
Comparing the lower limit \texttt{flux$\_$aper90$\_$lolim$\_$b} to the limiting sensitivity, 
the conservative selection reduces to 2552 candidates, so the model is still not excluded by the data, and the aggressive selection falls to 247 sources.

Finally, the uncertainties on the \Chandra~source flux can also affect the spectral constraints derived from the catalog. For instance, based on the Monte Carlo simulation, $\beta_H$ has to be less than one. This can be easily understood as the $H$ band ($2-7$ keV) is a sub-band of the broad band $B$ ($0.5-7$ keV). However, in the catalog, due to large flux error bars, ill-constrained sources can have $\beta_H$ larger than one. Relaxing the spectral cuts constraints to include these sources increases the number of candidates in the conservative selection, but the aggressive selection remains the same because its candidates are required to have a good spectral determination in order to meet its criteria. This should also be true for the other spectral constraints.

\begin{table*}[t]
    \setlength{\tabcolsep}{10pt}
    \centering
    \caption{Summary of systematic uncertainties explored in this work. We stress that the systematics explored in some cases represent extreme variations, and, as such, provide a very conservative estimate of the corresponding uncertainties. Unrealistic cases which overestimate the number of detectable MSPs are highlighted in \textit{italic}. 
    }
        \begin{tabular}{c|c|c|c|c}
        \hline\hline
     & Systematic & Test & Detectable MSPs & Aggressive candidates \\[2pt]\hline
    Baseline & --- & --- & \textbf{95$\pm$9}&\textbf{285}\\  \hline
    \multirow{7}{*}{Model}& \multirow{3}{*}{GLF} & $\alpha_1$ & 101$\pm$10&332\\  
    && $\alpha_2$ & 84$\pm$9&323\\  
    && $L_b$ & 78$\pm$9&241\\  \cline{2-5}
    & \multirow{2}{*}{$\gamma$-to-X} & multi-variate normal & 69$\pm$8&181\\  
    && {\it uncorrelated} & {\it 122$\pm$12} & {\it 353}\\ \cline{2-5}
    & \multirow{2}{*}{$N_H$} & {\it no absorption} & {\it 267$\pm$15} & 
    {\it 41} \\  
    && 2D & 75$\pm$8&368\\
    \hline
    \multirow{3}{*}{Observations}& \multirow{3}{*}{Sensitivity} & \texttt{TRUE} only & 54$\pm$7& 234\\  
    &&  flux upper limit & unchanged &303\\     
    &&  flux lower limit & unchanged &247\\     \hline\hline
\end{tabular}

\label{tab:sys_summary}
\end{table*}

\section{Prospects and conclusions}
\label{sec:concl}
We have shown {\it for the first time} that a simple
model for a population of MSPs in the Galactic bulge, which
can account for the excess $\gamma$-ray emission seen by the \Fermi-LAT, i) is consistent with current X-ray \Chandra~observations of compact sources and ii) together with information from \Gaia, it allows one to select the most promising few hundred \Chandra~sources for follow-up studies. On-going and future \Chandra\ observation programs covering our ROI will also increase the expected number of detectable MSPs, as well as candidate sources. 
Our work represents a first proof-of-principle of the 
potential of X-ray searches for bulge MSPs. Let us conclude by briefly discussing possible improvements as well as promising extensions of this analysis.

Potentially constraining information
is encoded in the spatial distribution of sources. This
can be exploited to 
optimize the ROI by maximizing the 
ratio of
detectable-to-candidate sources, see Appendix~\ref{app:StoN} for an illustration. 
However, such a procedure would strongly rely on the
assumption that the MSP spatial modeling is valid down to small scales, while we know that the $\gamma$-ray
observations on which it is based on are much more coarse grained. Therefore, we decide not to pursue our analysis further in this direction.

Our work rely on the $\gamma$-to-X correlation modeled in Sec.~\ref{sec:msppop}, and based on the observation of 40 MSPs having both X- and $\gamma$-ray detected emissions.
Characterizing the non-thermal multi-wavelength spectrum of a larger sample of MSPs (similarly to what done in~\cite{2020MNRAS.492.1025C,2019MNRAS.489.5494T} mostly for pulsars) 
would set population studies of MSP emission mechanisms on more solid grounds and improve 
our understanding of the $\gamma$-X connection.
Dedicated analyses of archived X-ray observations and existing source catalogs is another path to refine the spectral selection of MSP candidates, 
distinguishing them from other known population of faint X-ray sources. In our ROI, we expect numerous persistent Galactic sources such as chromospherically active stars, cataclysmic variables (CVs) and quiescent low-mass X-ray binaries. Most of foreground sources should already be excluded from our selections thanks to their \textit{Gaia} counterpart. More distant and partly highly absorbed bulge populations should be significantly different from the detectable MSP population described in Sec.~\ref{sec:det_MSP} but we cannot exclude that several of them could be included in our aggressive selection. The most critical overlap may be for CVs, and in particular intermediate polars (IPs), which have hard X-ray spectra, relatively high X-ray luminosities but faint optical counterparts compared with non-magnetic CVs, and are expected to be one of the dominant population in our ROI~\cite{2009ApJ...706..223H, 2011ApJS..194...18J}. The prominent iron emission lines observed in IPs may help to distinguish them from putative MSPs. In addition to Galactic populations, extragalactic sources may also contaminate our sample. In our ROI, the \textit{Chandra} catalog could contain up to $\sim120$ extragalactic sources with an observed 2--10 keV flux above $F_{\rm 2-10keV} = 2\times10^{-14}\rm \,erg/cm^{2}/s$~\cite{2005ApJ...635..214E}, which corresponds to both the average flux of our detectable MSP population and the mean ACIS sensitivity in our ROI. With an average photon index $\Gamma\sim1.7$, these extragalactic sources are therefore likely contaminating the faintest part of both our selections. These overlaps with the known populations of X-ray sources likely explain why our conservative selection exceeds by far the anticipated number of detectable MSPs. Constraining their precise contribution to our selections using multi-wavelength catalogs will therefore be key to further test our model.

A perhaps more promising extension of this study would be to engage in a multi-wavelength analysis of existing data, in particular in the radio and infrared bands, and to design follow-up campaigns to further isolate MSP candidates in the Galactic bulge~\cite{2012MNRAS.426.3057M} (see~\cite{2018MNRAS.479.2834H,2020ApJ...888L..18D} for applications to globular clusters).
Spectral characterization of MSPs vs.\ alternative sources are a pre-requisite for such studies. The sample selected with our ``aggressive'' cuts is the most interesting starting point for these further analyses that we plan to perform.

In conclusion, our findings open up 
new and exciting avenues to look for 
bulge MSPs and their connection with 
the GeV excess with X-ray observations. If supplemented with multiwavelength observations, these 
have the potential to provide breakthrough results in the near future.

\medskip
\paragraph{{\bf Acknowledgments.}}
We warmly acknowledge enlightening discussions with and comments on the draft by L. Guillemot.
We thank T.~D.~P.~Edwards for fruitful discussions about 
the $\gamma$-ray luminosity function. 
The PhD fellowship of J.~B.~is supported by the Mission pour les initiatives transverses et interdisciplinaires (MITI), CNRS (SMilERX project).
We acknowledge support from Agence Nationale de la Recherche AAPG2019, project GECO (PI: F.C.).
This work was supported by the Programme National des Hautes Energies of CNRS/INSU with INP and IN2P3, co-funded by CEA and CNES.
This research has made use of data obtained from the Chandra Source Catalog, provided by the Chandra X-ray Center (CXC) as part of the Chandra Data Archive and of data from the European Space Agency (ESA) mission Gaia, processed by the Gaia Data Processing and Analysis Consortium (DPAC).

\appendix

\section{Details of the Galactic MSP population}
\label{app:MSP_pop}

\paragraph{{\bf The MSP disk population.}} 
 Disk MSPs may represent an important background for MSP bulge searches. For the disk population, we use a ``Lorimer-disk'' profile~\cite{Lorimer06} whose number density is given by:
\begin{equation}
    \label{disk_number_density}
    \begin{aligned}
    n(r,z)= &\frac{NC^{B+2}}{4\pi R_\odot^2 z_s e^C \Gamma (B+2)} \left(\frac{r}{R_\odot}\right)^B \times\\ &\exp{\left(-C\left(\frac{r-R_\odot}{R_\odot}\right)\right)} \exp{\left(-\frac{|z|}{z_s}\right)} \, ,
    \end{aligned}
\end{equation}
with best-fit parameters $B = 3.91$, $C = 7.54$, defining the vertical 
and radial profile, and $z_s = 0.76$ pc the scale height~\cite{Bartels_mspluminosity}. 
$N$ is the normalisation of the source density distribution, which is set by the total number of sources, while $R_\odot$ is the Solar distance from the Galactic center, 
set to 8.5 kpc \cite{1986MNRAS.221.1023K}.
The best model for the ($0.1-100$ GeV) $\gamma$-ray luminosity function (GLF) of disk MSPs was found to be a broken power law \cite{Bartels_mspluminosity}:
\begin{equation}
   \frac{dN}{dL} \propto \left\{
    \begin{array}{ll}
        L^{-\alpha_1} & L \leq L_b\\
        L_b^{\alpha_2-\alpha_1} L^{-\alpha_2} & L > L_b
    \end{array}
    \right.
\label{eq:GLF}
\end{equation}
with $\alpha_1=0.97$, $\alpha_2=2.60$ and $L_b=10^{33.24}$ erg/s. 
\medskip

\paragraph{{\bf The MSP bulge population.}}
The MSP bulge population is made of two components:
The boxy bulge (BB) and the nuclear bulge (NB). 
The BB number density is proportional to $K_0(r_s)$ with $K_0$ being the modified Bessel function of the second kind, with $r_s$ given by:
\begin{equation*}
    r_s = \left[ \left( \left(\frac{x}{x_0}\right)^2 + \left(\frac{y}{y_0}\right)^2 \right)^2 + \left(\frac{z}{z_0}\right)^4 \right]^{\frac{1}{4}} \, ,
\end{equation*}
and with $x_0=$ 0.69 kpc, $y_0=$ 0.29 kpc and $z_0=$ 0.27 kpc \cite{Cao_bar_models}. Here, $(x,y,z)$ refer to the Cartesian BB 
coordinates system. The $z$ axis is perpendicular to the Galactic plane and the $x$ axis is rotated $\theta=29.4\degree$ away from the Galactic center-Sun axis in the clockwise direction \cite{Cao_bar_models}. \\
The NB \cite{Launhardt02}, in turn, gets contributions from 
the nuclear stellar cluster (NSC) and the nuclear stellar disk (NSD). For the NSD, the mass density in cylindrical coordinates is given by:
\begin{equation}
\begin{split}
   &\rho^{\rm NSD}(r,z) = \\
    &\left\{
    \begin{array}{ll}
        \rho^{\rm NSD}_{0} \left(\frac{r}{1 \, \rm pc}\right)^{-0.1}e^{-\frac{|z|}{45 \, \rm pc}} & r \leq 120  \, \rm  pc \\
        \rho^{\rm NSD}_{1} \left(\frac{r}{1 \, \rm pc}\right)^{-3.5}e^{-\frac{|z|}{45  \, \rm pc}} & 120  \, \rm pc < r \leq 220  \, \rm pc \\
        \rho^{\rm NSD}_{2} \left(\frac{r}{1 \, \rm pc}\right)^{-10}e^{-\frac{|z|}{45 \, \rm pc}} & r > 220\, \rm pc \\
    \end{array}
    \right.
\end{split}
\end{equation}
with $\rho^{\rm NSD}_{0}=301 \, M_{\odot} \, \rm pc^{-3}$ such that the mass within 120 pc is $8\times 10^8 \, M_{\odot}$. $\rho^{\rm NSD}_{1}$ and $\rho^{\rm NSD}_{2}$ are determined such to give a continuous NSD mass profile.
For the NSC, the mass density in spherical coordinates is given by:
\begin{equation}
   \rho^{\rm NSC}(r) = \left\{
    \begin{array}{ll}
        \frac{\rho^{\rm NSC}_{0}}{1+\left(\frac{r}{r_0}\right)^2} & r \leq 6 \, \rm pc \\[15pt]
        \frac{\rho^{\rm NSC}_{1}}{1+\left(\frac{r}{r_0}\right)^3} & 6 \, \rm pc < r \leq 200 \, \rm pc \\[15pt]
        0 & r > 200 \, \rm pc \\
    \end{array}
    \right.
\end{equation}
with $r_0$ = 0.22 pc and $\rho^{\rm NSC}_{0}=3.3\times10^6 \, M_{\odot}\, \rm pc^{-3}$. $\rho^{\rm NSC}_{1}$ is determined such to give a continuous NSC mass profile.

\medskip

\section{S/N spatial optimization}
\label{app:StoN}
\begin{figure*}[t!]
\includegraphics[width=1.\textwidth]{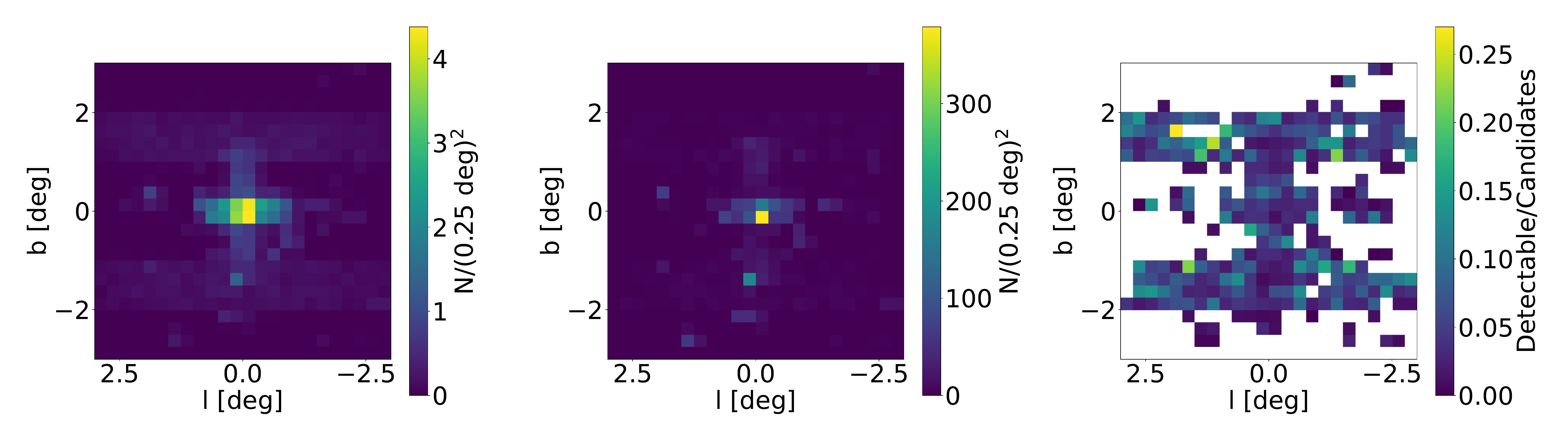}\\
\caption{{\bf Left panel}: 2D ($l$, $b$) histogram of detectable MSPs, 
averaged over 100 Monte Carlo simulations.
{\bf Central panel}: Same as left panel for the 3153 objects of the conservative selection. 
{\bf Right panel}: Detectable-to-candidate MSPs ratio.
}  
\label{fig:SNratios}
\end{figure*}

By exploiting the source spatial distribution, one could in principle improve the constraining power of the analysis. 
The synthetic bulge MSP population
has a specific distribution in space which traces 
the BB and the NB. Footprints of such a distribution 
are left in the $l$ and $b$ profiles of detectable MSPs.
In Fig.~\ref{fig:SNratios} (left panel), we show 
the 2D ($l$, $b$) histogram of detectable MSPs, where a 
clear ``cusp'' about the Galactic center direction can be seen. 
On the other hand, we do expect CSC selected sources not
to strictly follow the same distribution, given
the contamination from other source classes, see Fig.~\ref{fig:SNratios} (central panel).
Ideally, maximizing the ratio of detectable-to-candidate MSPs (Fig.~\ref{fig:SNratios}, right panel) one can optimize the ROI and design a 
strategy to further cut down the candidates' sample.
As discussed in the main text, this approach would however require 
to model the MSP spatial distribution on small scales, a goal which cannot be reliably achieved based on current $\gamma$-ray analyses and/or theoretical models.

\bibliography{mspxray}

\end{document}